\documentclass[runningheads]{llncs}
\usepackage{graphicx}
\usepackage{changepage}   
\usepackage[hidelinks]{hyperref}

\usepackage{tabularx}
\usepackage{caption} 
\newcommand{\header}[1]{\textit{#1}}

\begin{document}
\title{Content negotiation on the Web: State of the art}

\author{Yousouf Taghzouti\inst{1}\orcidID{0000-0003-4509-9537}\and\\
Antoine Zimmermann\inst{1}\orcidID{0000-0003-1502-6986} \and\\
Maxime Lefrançois\inst{1}\orcidID{0000-0001-9814-8991}}
\authorrunning{Y. Taghzouti et al.}
%
\institute{Mines Saint-Etienne, Univ Clermont Auvergne, INP Clermont Auvergne, CNRS, UMR 6158 LIMOS, F - 42023 Saint-Etienne France\\
\email{\{yousouf.taghzouti,antoine.zimmermann,maxime.lefrancois\}@emse.fr}}
\maketitle              
\begin{abstract}
The openness and accessibility of the Web has contributed greatly to its worldwide adoption. Uniform Resource Identifiers (URIs) are used for resource identification on the Web. A resource on the Web can be described in many ways, which makes it difficult for a user to find an adequate representation. This situation has motivated fruitful research on content negotiation to satisfy user requirements efficiently and effectively. We focus on the important topic of content negotiation, and our goal is to present the first comprehensive state of the art. Our contributions include (1) identifying the characteristics of content negotiation scenarios (styles, dimensions, and means of conveying constraints), (2) comparing and classifying existing contributions, (3) identifying use cases that the current state of content negotiation struggles to address, (4) suggesting research directions for future work. The results of the state of the art show that the problem of content negotiation is relevant and far from being solved.

\keywords{Content negotiation \and Style of negotiation \and Dimension of negotiation  \and Constraints.}
\end{abstract}
%
%
\section{Introduction}

Open, distributed, accessible, and heterogeneous are some of the fundamental characteristics of the Web~\cite{choudhury:2014}. Although the fact that anyone can access the Web from anywhere in the world has greatly contributed to its development and enrichment through its openness, this has had the opposite effect of having an abundance of Web resources and difficulties in providing the best content for each client; a simple example is that of two people speaking different languages accessing the same resource, the server with the resource should be able to provide each client with an understandable version. To remedy this, a solution was devised from the start, with a negotiation layer between the client and the server~\cite{berners-lee_cailliau_groff_pollermann:1992}. It is discussed later in the Architecture of the World Wide Web document as one of the essential components of Web design~\cite[section~3.2]{web-arch_w3c:2004}.
     
Negotiation, as a concept, is a back-and-forth communication intended to reach an agreement when two or more parties have common and opposing interests~\cite[p.~1]{fisher:2011}. Applied to the Web, it then becomes ``content negotiation'' (CN) the mechanism for selecting the appropriate representation when processing a request. In HTTP, constraints called preferences can be expressed and transmitted as mentioned in~\cite[section~5.3]{http_rfc:2014}. And with this, in addition to finding and transmitting information, it is possible to select more specific formats and languages. 

With the mobile era and network constraints came a new challenge, the content already available was designed to fit computer screens, not phones. Once again, content had to be negotiated in order to know what was appropriate for these devices based on their characteristics~\cite{cc_pp_w3c:2004}.

\subsection{Problem statement}
Resources on the Web are accessible via uniform resource identifiers (URIs). One resource may have different alternatives that we call variants as in~\cite{http_rfc:1999}. To request a specific variant a client would use CN. Thereby, CN is the mechanism of choosing the best variant from among a set of alternatives of a resource available on the Web under a URI. The request sent by a client contains a set of constraints that would enable the server to provide the adequate response. The server's handling of the request may vary in accordance with the used CN style and CN dimension as well as the constraint conveyance technique and the employed protocol. 

\subsection{Methodology and scope of survey}
We used the following strategy to identify the content negotiation features as well as the available contributions that are the subject of our classification in our state of the art.
We used the keyword ``content negotiation'' to search for papers in Google Scholar\footnote{\url{https://scholar.google.com/}}, The DBLP Computer Science Bibliography\footnote{\url{https://dblp.org/}} and Semantic Scholar\footnote{\url{https://www.semanticscholar.org/}}.
We chose relevant articles, reports empirical study from the collected resources based on whether their studied problem is compatible with our problem statement presented in Section 1.1. Furthermore, we followed citations and references to consider additional resources that cite or are cited by the above relevant resources.

\subsection{Related problems and surveys}
Efforts were done to encapsulate the available techniques and characteristics of CN, but to the best of our knowledge this only took the form of a related work section in specifications using CN such as~\cite[Section 5]{conneg_profile_w3c:2019} or web page documentation such as in Mozilla Developer Network Web Docs\footnote{\url{https://developer.mozilla.org/en-US/docs/Web/HTTP/Content_negotiation}}.

Another related problem is personalised information retrieval (PIR) that aims not only to assist users in finding information from the myriad of information resources available on the Web as in traditional information retrieval field, but takes into account user preference and the history of their interactions with the system with the main objective to increase user satisfaction. A survey was conducted and proposed a classification of PIR systems~\cite{ghorab:2013}.

\subsection{Contribution of the article}
The broad application of content negotiation has yielded fruitful research results in recent years. However, a comprehensive state of the art on this research topic does not exist. To the best of our knowledge, we provide the first state of the art of existing content negotiation approaches and features. We identify the characteristics of each contribution and categorise them in order to classify the existing contributions on content negotiation. Based on this state of the art, we propose use cases that are not yet well addressed and suggest directions for future work.

\subsection{Structure of the article}
The remainder of this article is organised as follows. Section~\ref{sec:term} presents the terminology of CN. Section~\ref{sec:characteristics} reviews the characteristics of CN and groups them into categories, allowing us in Section~\ref{sec:benchmark} to provide a comparative analysis of a list of contributions in this area. Finally, we conclude with the presentation of two use cases that CN in its current state is not able to fully address and suggest a future direction in Section~\ref{ref:conclusion}.

\section{Terminology}
\label{sec:term}
This section introduces the basic terminology of content negotiation:
\begin{description}
    \item[Capability] is an attribute of a sender or receiver (often the receiver) which indicates an ability to generate or process a particular type of message content~\cite{rfc2703:1999}.
    \item[Client] is a program that establishes connections for the purpose of sending requests~\cite{http_rfc:1999}.
    \item[Negotiable resource] is a data resource which has multiple representations (variants) associated with it. Selection of an appropriate variant for transmission in a message is accomplished by content negotiation between the sender and recipient~\cite{rfc2703:1999}.
    \item[Negotiation dimension] also called \textit{constraints} or \textit{preferences:} indicate the constraints considered when selecting the best representation.
    \item[Negotiation style] is a particular way or technique by which the CN process is conducted. This includes how the negotiation is conducted and which party of the CN chooses the variant to be selected.
    \item[Resource] is a network data object or service that can be identified by a URI. Resources may be available in multiple representations (e.g. multiple languages, data formats, size, and resolutions) or vary in other ways~\cite{http_rfc:1999}.
    \item[Server] is an application program that accepts connections in order to service requests by sending back responses. Any given program may be capable of being both a client and a server; our use of these terms refers only to the role being performed by the program for a particular connection, rather than to the program's capabilities in general. Likewise, any server may act as an origin server, proxy, gateway, or tunnel, switching behaviour based on the nature of each request~\cite{http_rfc:1999}.
    \item[User-Agent] is the client which initiates a request. These are often browsers, editors, spiders (web-traversing robots)~\cite{http_rfc:1999}.
    \item[Variant] is a representation that is associated with a resource at a given time~\cite{http_rfc:1999}.
\end{description}

\section{Content negotiation characteristics}
\label{sec:characteristics}
We broadly divide the characteristics of CN into \textit{dimensions}, \textit{styles}, and \textit{means of constraint conveyance} which we will describe in the next subsections. 

\subsection{Content negotiation dimension}
A content negotiation dimension also known as \textit{constraint} or \textit{preference} that denotes the constraints the server took into consideration when selecting the best representation. CN dimensions are the primary differentiating factor between a set of alternatives. We say that two alternative representations vary according to one or multiple CN dimension.

\subsubsection{Accuracy}
Measurement is a process that uses numbers to describe a physical quantity. Units of measurement provide standards so that the numbers in our measurements refer to the same thing. A user may want to request a representation that meets a particular precision or unit of measurement, e.g. ucum~\cite{schadow:2005}.

\subsubsection{Authorisation}
Data on the Web like our behaviours in the real world, could be subject to regulation. Resources could be restricted, for example, for privacy reasons. Users may want to request data only if it complies with certain regulation and if they have the proper permissions.

\subsubsection{Encoding}
In the HTTP protocol, the user can express a preference for a compression algorithm using the \header{accept-encoding} header~\cite[Section 5.3.4]{http_rfc:2014}. The accepted values are: \textit{gzip}, \textit{compress}, \textit{deflate}, \textit{br}, \textit{identity} or \textit{*}.

\subsubsection{Capability}
In general, an attribute that defines the capabilities and preferences of a receiver's hardware or software\footnote{An example of a Nokia CC/PP file \url{http://nds1.nds.nokia.com/uaprof/N6230ir200.xml}}, e.g., color: the receiver can process color, grayscale, or only black and white, screen size: height and width, etc~\cite{butler:2001}.

\subsubsection{Char-Set}
Before \textit{UTF-8} was widely supported, the user could negotiate the encoding of the characters it supports, but this is no longer the case. In HTTP, the user expressed this preference using the \header{Accept-Charset} header~\cite[Section 5.3.3]{http_rfc:2014}.

\subsubsection{Coordinate Reference Systems (CRS)}
Objects on the earth can be located using a coordinate reference system (CRS). Different users may be interested in a representation of these objects in different systems: a local, regional or global system, for example WGS84 or ETRS89\footnote{https://github.com/opengeospatial/conneg-by-crs}.

\subsubsection{Entailment regime}
Entailment regimes allows additional RDF statements to be inferred from explicitly given assertions. A user may want to request data that conforms to a form but after using a specific entailment regime such as RDF Schema~\cite{glimm:2013}.

\subsubsection{Formatting}
Formatting in general refers to the way text, images, etc. are organised. A user may want to request a representation that respects a given formatting, such as the minified version of a JS file or the citation style of a bibliography.

\subsubsection{Language}
In the HTTP protocol, the user can express a preference for languages using the \header{accept-language} header~\cite[Section 5.3.5]{http_rfc:2014}.

\subsubsection{Media type}
In the HTTP protocol, the user can express a preference for media types\footnote{List of media types: \url{https://www.iana.org/assignments/media-types/media-types.xhtml}} using the \header{accept} header~\cite[Section 5.3.2]{http_rfc:2014}.

\subsubsection{OWL Profile}
An OWL profile is a reduced version of \textit{OWL Full} that trades some expressive power for reasoning efficiency. For this reason, users may wish to request an ontology in a specific OWL profile such as \textit{OWL~2 EL}~\cite{owl_w3c:2012}.

\subsubsection{Summary}
The abundance and length of representations on the Web make them increasingly difficult for humans to use. A user may want to request a summary of a representation with specific characteristics e.g. word number or number of paragraphs, instead of receiving the full original representation.

\subsubsection{Time}
A resource could have different representations of the state that it went through in time. \cite{rfc7089:2013} introduced a framework to negotiate the resource states through datetime negotiation, by adding the headers \header{Accept-Datetime}.

\subsubsection{Vocabulary}
RDF is a general language for representing information on the Web. This information on the Web can be described using different vocabularies. Users may want to request representations described using a certain vocabulary, for example data about a person described by FOAF or vCard.

\subsubsection{The profile dimension}
With the dimensions mentioned above, a line can be drawn between them (they are somehow orthogonal), but it is worth mentioning the profile dimension introduced in~\cite{atkinson_w3c:2019} to allow content negotiation by profile~\cite{conneg_profile_w3c:2019}. A profile can have \textit{Resource Descriptors} associated with it, each resource descriptor must indicate the role it plays, among others: constraints, orientation, vocabulary. This implies that a profile dimension encapsulates sub-dimensions and that the implementer must check which role to apply in the negotiation process. 

\subsection{Content negotiation style}
We use CN style to denote a particular manner or technique by which the content negotiation process is carried out. This includes how the negotiation is conducted and which of the content negotiation parties makes the choice of which variant to select. In our study, we identified six CN styles: \textit{proactive}, \textit{reactive}, \textit{transparent}, \textit{active}, \textit{conditional}, \textit{adaptive}. These styles are not mutually exclusive, and each has trade-offs in applicability and practicality. In the following subsections, a brief description of each style is provided, along with some generic advantages and disadvantages associated with its use.
\subsubsection{Proactive}
When content negotiation preferences are sent by the user agent in a request to encourage an algorithm located at the server to select the preferred representation, it is called proactive negotiation (or server-driven negotiation). Selection is based on the available representations for a response (the dimensions over which it might vary, such as language, content-coding, etc.) compared to various information supplied in the request~\cite{http_rfc:2014}.
\paragraph{Advantages}
\begin{itemize}
    \item The server avoids the back and forth because the client sends the preferences to the server which makes its best guess and sends them with the answer.
    \item The sever does not need to describe the selection algorithm to the client to make a choice.
\end{itemize}
\paragraph{Disadvantages}
\begin{itemize}
    \item It is impossible for the server to accurately determine what might be ``best'' for any given user, since that would require complete knowledge of both the capabilities of the user agent and the intended use for the response.
    \item Having the user agent describe its capabilities in every request can be both very inefficient (given that only a small percentage of responses have multiple representations) and could be a potential risk to the user's privacy.
    \item It complicates the implementation of an origin server and the algorithms for generating responses to a request.
    \item It limits the reusability of responses for shared caching.
\end{itemize}
\subsubsection{Reactive}
If the server receives an ambiguous request, it sends a list of the different alternatives it has. The user agent can make the choice if it has sufficient knowledge of the user's preferences. Otherwise, it displays the list of links for the user to make the final choice~\cite{http_rfc:2014}.
\paragraph{Advantages}
\begin{itemize}
    \item The cache is used to reduce network overhead.
    \item The client gains more privacy because no description is sent to the server.
\end{itemize}
\paragraph{Disadvantages}
\begin{itemize}
    \item The increase of latency due to the round trip to select the representation.
\end{itemize}
\subsubsection{Transparent}
Transparent content negotiation is called ``transparent'' because it makes visible to the intermediate parties (between the origin server and the user agent e.g. proxy cache) all the variants that exist within the origin server and gives them the ability to choose the best representation on their behalf. Transparent content negotiation is a combination of proactive and reactive content negotiation. In reactive content negotiation, when a cache is provided in the form of a list of available representations of the response and the intermediate party has fully understood the dimensions of the variance, then the intermediate party becomes capable of performing proactive content negotiation on behalf of the origin server for subsequent requests on that resource~\cite{conneg_rfc:1998,seshan:1998,http_rfc:2014}.

\paragraph{Advantages}
\begin{itemize}
    \item The reduction of response time and bandwidth consumption due to the distribution of the negotiation work that would otherwise be required of the origin server.
    \item The gain of the second request delay of reactive negotiation when intermediary party uses the cache to be able to correctly guess the right response.
\end{itemize}

\paragraph{Disadvantages}
\begin{itemize}
    \item The underlined assumption of maximum resource cachability which in practice is only true for static and unencrypted content, implying that it would not be cost effective in contexts where transmissions are ciphered.
\end{itemize}

\subsubsection{Conditional}
The response that a server replies with to a request in a conditional content negotiation style consists of a body composed of several parts separated by boundaries. The parts are selectively rendered based on the user agent's parameters. 
This can take the form of parts containing different variants of the resource, for example with distinct media types\footnote{\url{https://docs.marklogic.com/9.0/guide/rest-dev/bulk}}, or parts containing portions of a representation, for example certain pages of a PDF document~\cite{fielding:2014,http_rfc:2014}.

\paragraph{Advantages}
\begin{itemize}
    \item The reduction of the number of queries to a single request that gets a multi-part response containing multiple variants.
    \item The ability to select only a portion of a representation. 
\end{itemize}

\paragraph{Disadvantages}
\begin{itemize}
    \item The nonscalability of this style if the number of variants or the size of variants is large.
\end{itemize}

\subsubsection{Active}
The server in the active CN responds with a reply that contains a script. The script makes additional (more specific) requests based on the characteristics of the user agent~\cite{http_rfc:2014}.
\paragraph{Advantages}
\begin{itemize}
    \item  The reduction of user interaction by automating the sending of additional requests.
    \item The provision of a personalised representation that matches the capabilities of the user agent.
\end{itemize}

\paragraph{Disadvantages}
\begin{itemize}
    \item The need for multiple requests to build the final representation. 
    \item The introduction of potential threats due to script execution, e.g. a man-in-the-middle attacker can intercept or rewrite the response to include malicious JavaScript code. Malicious active content can steal the user's credentials, acquire sensitive data about the user, or attempt to install malware on the user's system (by leveraging vulnerabilities in the browser or its plugins, for example).
    \item The prohibition of active content by default in the most recent versions of browsers due to the vulnerabilities mentioned above.
\end{itemize}

\subsubsection{Adaptive}
The server in adaptive CN replies with a representation that has most likely undergone an adaptation process. Adaptation can be performed internally by the server or by using another service to perform this task, for example: performing a transformation that requires a lot of processing power~\cite{lemlouma:2002}. Adaptation is successful if the final delivered representation is more understandable to the user based on his/her context.

\paragraph{Advantages}
\begin{itemize}
    \item Increase the satisfaction rate of the constraints due to additional adaptation.
\end{itemize}

\paragraph{Disadvantages}
\begin{itemize}
    \item The nonscalable nature of the pre-adaptation process of content.
    \item The invasive nature of describing a client context.
\end{itemize}

\subsection{Conneg constraint conveyance}
In a content negotiation process, the client must transmit to the server the negotiation dimension as well as its value to be considered for the process of selecting the best variant to provide. Two main techniques have emerged and are widely used to perform the transmission of constraints: The \textit{HTTP headers} and the \textit{URI} based approaches. The following subsections describe each of these two techniques.

\subsubsection{HTTP headers}
HTTP headers are essential elements of the HTTP protocol that allow additional information to be transmitted by the client (request headers) and the server (response headers). This section will show how some HTTP headers are leveraged to convey constraints. It is worth mentioning that CoAP also implements CN, but only for formats via the \header{accept} option~\cite[Section 5.10.4]{coap_rfc:2014}.

\paragraph{Accept:} a request header that indicates which media type expressed in MIME types the client prefers. The server selects a variant and informs the client of its choice with the \header{Content-Type} response headers.

\paragraph{Accept-Language:} a request header that indicates the language expressed by the natural language and locale that the client prefers. The server selects a variant and informs the client of its choice with the \header{Content-Language} response headers.

\paragraph{Accept-encoding:} a request header that indicates the encoding typically a compression algorithm that the client prefers. The server selects a proposal and informs the client of its choice with the \header{Content-Encoding} response headers.

\paragraph{Accept-Crs:} a request header that indicates the CRS that the client prefers. The server selects a proposal and informs the client of its choice with the \header{Content-Crs} response headers\footnote{\url{https://github.com/opengeospatial/conneg-by-crs/}}.

\paragraph{Accept-Presentation:} a request header that indicates the presentation that the client prefers. The server selects a proposal and informs the client of its choice with the \header{Content-Presentation} response headers~\cite{lefrancois:2018}.

\subsubsection{URI based}
URIs not only provide a simple and extensible way to identify resources on the Web, they could be used to convey constraints to guide the selection of a preferred variant. In this section, we present three ways to utilise them. 

\paragraph{Archival Resource Key (ARK)} ARK is an identification scheme for a persistent identifier for information objects. Using the optional ``Qualifier'' part, a kind of service entry point could be created that allows an ARK to support access to variants (versions, languages, formats) of components by using the '.' (period) character after the Name part of an ARK. For example, the URI \footnote{\url{https://api.istex.fr/ark:/67375/6GQ-MLC8GRWC-5}} list all the variants and the PDF variant is at the URI \footnote{\url{https://api.istex.fr/ark:/67375/6GQ-MLC8GRWC-5/fulltext.pdf}}~\cite{viot:2018}.

\paragraph{URI path extension (suffix pattern matching)} this approach is similar to the ARK approach, which is to use the URI with an extension primarily to request the media type of API endpoints. The URI\footnote{\url{http://myapi.example.com/account/123.json}} would provide the 123 account in a json format. 

\paragraph{Query String Arguments (QSA)} A query string is a part of a URI which assigns values to specified parameters. QSAs are commonly used to deliver extra information to a server. One of which is preferences to select an appropriate variant\footnote{\url{http://linked.data.gov.au/dataset/gnaf/address/GAACT714845933?_view=ISO19160&_format=text/turtle}}.

\section{Existing contribution benchmark}
\label{sec:benchmark}
The table \ref{tab:contribution-benchmark} presents our effort to gather some known contributions in the literature pushing the state of the art in different directions but having in common their use of CN, each contribution is presented with a reference and its publication date as well as its respective characteristics according to our previously presented categorisation.
   
\begin{table*}[t]
    \centering    
    \begin{tabularx}{1\textwidth}{ p{0.75cm} p{1cm} p{3.5cm} p{2.5cm} p{2cm} p{3cm} }
      \hline
     Ref & Date & Style & Dimension & Transmission  & Protocol\\ 
     \hline
     \cite{butler:2001} & 2001 & Proactive & Capacity & header & HTTP\\ 
     \cite{lum:2003}& 2003 & Adaptive & Capacity & QSA & HTTP\\ 
     \cite{viot:2018}& 2018 & Reactive & Media type & ARK & HTTP\\ 
     \cite{kelly:2018}& 2018 & Reactive & Time & header & HTTP\\ 
     \cite{lefrancois:2018} & 2018 & Proactive, Adaptive & Presentation & header & HTTP, CoAP\\ 
     \cite{markLogic:2018}& 2018 & Conditional & Media type & header & HTTP\\
     \cite{ocf:2022} & 2022 &  & Media type & header & CoAP\\ 
     
     \hline
    \end{tabularx}
    \caption{Contributions using CN (sorted by publication date) and their characteristics (Blank: not known).}
    \label{tab:contribution-benchmark}
\end{table*}

Composite Capabilities/Preferences Profile (CC/PP) and WAP UAProf are descriptions of device capabilities and user preferences. \cite{butler:2001} describes an implementation of HTTP content negotiation that uses them to provide the best variant. We believe that the proactive style of CN is used and, according to our characteristics, the dimension is \textit{capacity}, the headers are used to convey constraints, and the protocol is HTTP.

The paper~\cite{lum:2003} presents a decision engine capable of determining optimal adaptation decisions from interpolating situational contextual information, e.g., device capacity. HTTP was used by embedding the userid in the URI to identify the end user.

Istex is a French scientific archive~\cite{viot:2018}. The clients have the possibility to have the representation in several formats and for that, as mentioned on the site, an ARK is used. If we request the resource without specifying the media type, we receive a json file describing the existing variants and therefore we can consider it a reactive negotiation. The dimension is the media type and the transmission in ARK using the HTTP protocol.

Web archives play a major role in providing a picture reflecting the state of the Web at a given time. The contribution~\cite{rfc7089:2013} involves the client request in the Memento aggregation process beyond the specification of a URI-R and a date as described in~\cite{kelly:2018} using the header \header{prefer}. Therefore the characteristics are: time for dimension, headers for constraint transport, and HTTP as the protocol and because the aggregator returns a set of records and the client can potentially manipulate the response to issue another request, we judge this part to have a reactive CN style.

One scenario described in the contribution~\cite[Section 3.2]{lefrancois:2018} is of a client that requests a resource representation from a server and wants the response to be encoded according to a specific RDF presentation, which is done by including metadata in the request header. And because this contribution is primarily intended for constrained devices, the proactive + adaptive style is preferable\footnote{It is important to mention that the content negotiation styles are not mutually exclusive}. The author points out that, although this method has been implemented in HTTP, it could easily be defined as equivalent CoAP options.

Using the MarkLogic~\cite{markLogic:2018} server, a REST application developer uses the conditional CN style. A client would receive a response with a body containing several parts separated by a delimiter to be selected. This method is primarily used to select the media type dimension and uses HTTP and headers to convey this constraint. In addition, the Open Connectivity Foundation (OCF) specification also includes the means to negotiate the media type using the CoAP protocol~\cite[Section 12.2.4]{ocf:2022}. 

%
%
\section{Conclusion and Future work}
\label{ref:conclusion}
Content negotiation is a fundamental mechanism of the Web. In this state of the art, we have presented an analysis of the characteristics of existing CN scenarios, including styles, dimensions, means of constraint transmission, and CN protocols. These characteristics can be used to rank existing contributions as shown in the table~\ref{tab:contribution-benchmark}. 

However, some use cases do not yet have a satisfactory solution, e.g., the case of vocabulary negotiation where a client wants a way to search for representations in a specific vocabulary. Or to specify the desired vocabularies. For example, requesting that the creator's data use the FOAF (Friend Of A Friend), Schema.org or DCMI (Dublin Core Metadata Initiative) vocabulary. If the data graphs available on the server use the same media type, for example: \textit{text/turtle}, the client must manually query all data graphs to select those that use the desired vocabulary.

Another limitation is highlighted with the use case of RDF shape negotiation, where a client needs a representation that conforms to a specific shape, so even vocabulary negotiation is not sufficient because the client would have to manually validate all returned data graphs. In this case, negotiation can be rigid in the case where the client wants \textit{all} the constraints to be valid, and prefers not to have an answer otherwise. On the other hand, negotiation can be flexible in the case where the client agrees to receive a representation even if it does not satisfy all the constraints. And finally, these form constraints may not have the same degree of importance. Thus, the client may want a way to express this importance for each constraint and get the representation that accounts for it.

For future work, we will focus on creating an online resource (a website) that collects and categorises CN features following the structure proposed in this study: styles, dimensions, etc. This resource will also aim to collect CN use cases, highlight existing solutions if available, or suggest plausible ways to advance them. The goal will be to have an up-to-date digital survey of CN.

%
%

%
%
%
\bibliographystyle{splncs04}
\bibliography{biblio.bib}

\end{document}